\documentclass{ws-ijmpd}
\usepackage[super,compress]{cite}
\usepackage[pdftex,breaklinks]{hyperref}
\hypersetup{colorlinks,urlcolor=black,citecolor=black,linkcolor=black,filecolor=black}
\usepackage{breakurl}

\usepackage{amsfonts}
\usepackage{amssymb}
\usepackage{amsmath}
\usepackage{mathrsfs}
\usepackage{slashed}
\usepackage{textcomp}

\newcommand{\Tr}{\mathop{\mathrm{Tr}}}
\DeclareMathOperator{\sgn}{sgn}

\begin{document}

\markboth{V. BEYLIN, V. KUKSA, N. VOLCHANSKIY}
{Models of hypercolor based on symplectic gauge group \dots}

%
\catchline{}{}{}{}{}
%

\title{Models of hypercolor based on symplectic gauge group\\with three heavy vectorlike hyperquarks}

\author{NIKOLAY VOLCHANSKIY}

\address{Research Institute of Physics, Southern Federal University,\\
Prospekt Stachki 194, 344090 Rostov-na-Donu, Russia\\
nikolay.volchanskiy@gmail.com}

\address{Bogoliubov Laboratory of Theoretical Physics,\\
Joint Institute for Nuclear Research,\\
ulitsa Joliot-Curie 6, 141980 Dubna, Moscow region, Russia}

\author{VLADIMIR KUKSA}

\address{Research Institute of Physics, Southern Federal University,\\
Prospekt Stachki 194, 344090 Rostov-na-Donu, Russia\\
vkuksa47@mail.ru}

\author{VITALY BEYLIN}

\address{Research Institute of Physics, Southern Federal University,\\
Prospekt Stachki 194, 344090 Rostov-na-Donu, Russia\\
vitbeylin@gmail.com}
\maketitle

\begin{history}
\received{Day Month Year}
\revised{Day Month Year}
\end{history}

\begin{abstract}
We study possibilities to extend the Standard Model (SM) by three flavors of vectorlike heavy quarks in pseudoreal representation of symplectic hypercolor gauge group. This extension of SM predicts a rich spectra of heavy composite hypermesons and hyperbaryons (all of them carry integer spins) including fourteen pseudo-Nambu--Goldstone states emerging in dynamical breaking of the global symmetry group of the H-quarks, SU(6), to its Sp(6) subgroup. The properties of the lightest states  depend strongly on the choice of heavy-quark hypercharges. Our focus is placed on the variants of the model with partially composite Higgs boson, i.e.\ the experimentally observed boson is comprised of the elementary SM Higgs and a mixture of H-hadrons.
\end{abstract}


\ccode{PACS numbers: 12.60.Rc; 11.30.Er; 11.30.Fs; 11.15.Ex}



\section{Introduction}

Since the long-awaited discovery of the Higgs boson \cite{1H,2H} with a mass of 125 GeV in 2012, much experimental effort has been being directed toward establishing the properties of the boson, which, one may hope, might result  eventually in detecting indications of physics beyond the Standard Model (SM). Among the possibilities of new effects that could manifest itself in the Higgs sector are (1) the existence of additional states and (2) hypothetical composite nature of the Higgs boson (see Refs.~\refcite{2016LNP...913.....P,2014EPJC...74.2766B} for extensive reviews and bibliographies).

One of the most simple ways to introduce (partially) composite and optionally multi-state Higgs sector is to extend SM by a set of strongly interacting vectorlike fermions, hyperquarks (H-quarks), in confinement under a new H-color gauge group \cite{Sundrum,Pasechnik:2013bxa,Pasechnik:2013kya,Lebiedowicz:2013fta,Pasechnik:2014ida,doi:10.1142/S0217751X17500427,2015JHEP...12..031A,2017PhRvD..95c5019A,2018JHEP...08..017B,2015JHEP...01..157A,2017JHEP...10..210M}. Provided quantum numbers of H-quarks are suitably chosen, a composite Higgs doublet or doublets can be present in the spectrum of H-hadrons. Another option is to consider the SM Higgs as originally elementary state, but partially composite in view of its mixing with some of the H-hadrons. On the other hand, as the H-quarks are vectorlike (chirally symmetric) with respect to the symmetries of SM, such hypercolor models usually are in a good agreement with electroweak (EW) precision constraints.

The simplest realizations of the scenario described include, for example, models with two vectorlike H-flavors in confinement under H-color group SU(2) \cite{2017PhRvD..95c5019A,Beylin:2016kga}. In this work, however, we consider a model with richer symmetry structure. Our model comprises three H-flavors charged under a new symplectic H-color group $\text{Sp}(2\chi_{\tilde{c}})$. The representations of the group are pseudo-real, which implies that the global symmetry group of the new strong sector is SU(6) and larger than the chiral group of usual QCD. The global symmetry group SU(6) is broken spontaneously to Sp(6). As a consequence, fourteen pseudo-Nambu--Goldstone states are present in the model. Their number is enough to construct a composite two Higgs doublet model as is done in Ref.~\refcite{2018arXiv180507619C}. The same coset SU(6)/Sp(6) was also used for realizations of the little Higgs scenario \cite{Low:2002ws,Csaki:2003si,Gregoire:2003kr,Han:2005dz,Brown:2010ke,Gopalakrishna:2015dkt}. Here, we consider the case of partial compositeness---our model contains the usual elementary SM Higgs doublet that mixes with H-hadrons.

The remainder of the paper is organized as follows. The next section \ref{sec:lag/sym} discusses the Lagrangian and global symmetry group of the underlying ultraviolet completion under consideration. In section \ref{sec:LSM} an effective theory of H-quark and H-hadron interactions is constructed as linear sigma model based on the coset SU(6)/Sp(6). Section \ref{sec:vars} is devoted to a consideration of auxiliary symmetries of the model. The Peskin-Takeuchi parameters are calculated for some physically interesting scenarios in section \ref{sec:pt}. The results are summarized in the conclusion. The generators of SU(6) and Sp(6) groups are listed in the appendix.


\section{\label{sec:lag/sym}Lagrangian and global symmetry of symplectic QCD with 3 hyperquark flavors}

Let us consider an extension of SM with an additional symplectic H-color group, i.e.\ the gauge group of the model is $G=G_\text{SM} \times \text{Sp}(2\chi_{\tilde{c}})$, $\chi_{\tilde{c}} \geqslant 1$. Note that in the simplest case $\chi_{\tilde{c}} = 1$ we have an isomorphism $\text{Sp}(2)=\text{SU}(2)$. We introduce six left-handed Weyl hyperquarks grouped initially as two weak doublets $Q_{\text{L}(A)}^{k\underline{k}}$ and two singlets $S_{\text{L}(A)}^{\underline{k}}$, where underlined indices correspond to the fundamental representation of the hypercolor group $\text{Sp}(2\chi_{\tilde{c}})$ and $A=1,\,2$. The H-quarks fields transform under the gauge group $G$ as
\begin{align}\label{2.1}
(Q^{j\underline j}_{\text{L}(A)})'&{}=Q^{j\underline j}_{\text{L}(A)}-\frac{i}{2} g_1 Y_{Q(A)} \theta
Q^{j\underline j}_{\text{L}(A)}+\frac{i}{2}g_2 \theta_a
\tau_a^{jk}Q^{k\underline j}_{\text{L}(A)}+\frac{i}{2}g_{\tilde{c}}\theta_{\underline a}\lambda_{\underline a}^{\underline j \underline k}
Q^{j\underline k}_{\text{L}(A)},
\\ \label{2.2}
(S^{\underline j}_{\text{L}(A)})'&{}=S^{\underline j}_{\text{L}(A)}-ig_1 Y_{S(A)} \theta
S^{\underline j}_{\text{L}(A)}+\frac{i}{2}g_{\tilde{c}}\theta_{\underline a}\lambda_{\underline a}^{\underline j \underline k}S^{\underline k}_{\text{L}(A)}.
\end{align}
Here $\theta$, $\theta_a$, $\theta_{\underline{a}}$ are transformation parameters of $U(1)_\text{Y}$, $SU(2)_\text{L}$, and $\text{Sp}(2\chi_{\tilde{c}})_{\tilde{c}}$ respectively; $\lambda_{\underline a}$, $\underline{a} = 1 \dots \chi_{\tilde{c}} (2\chi_{\tilde{c}} + 1)$ are generators of the hypercolor group.  The generators satisfy the relation
\begin{align}\label{eq:scgr}
\lambda_{\underline a}^{\underline{k} \underline{j}} \varepsilon_{\tilde{c}}^{\underline{k} \underline{l}} + \varepsilon_{\tilde{c}}^{\underline{j} \underline{k}} \lambda^{\underline{k} \underline{l}}_{\underline a} = 0,
\end{align}
where $\varepsilon_{\tilde{c}}$ is an antisymmetric $\text{Sp}(2\chi_{\tilde{c}})$ ``metric'' normalized so that $\varepsilon_{\tilde{c}}^2 = -1$. In what follows, we will omit $SU(2)_\text{L}$ and $\text{Sp}(2\chi_{\tilde{c}})_{\tilde{c}}$ indicies.

The H-quarks transform as self-contragredient (pseudo-real) representations of the gauge group, i.e. the relation \eqref{eq:scgr} holds true. Note also that the analogous relation, $\tau_a^\text{T} \varepsilon + \varepsilon \tau_a =0$ ($\text{T}$ stands for ``transpose'') with $\varepsilon=i\tau_2$, is valid for the Pauli matrices, since $\text{Sp}(2)=\text{SU}(2)$. We can build up right-handed fields from left-handed ones
\begin{align}
    Q_{\text{R}(A)} = \varepsilon \varepsilon_{\tilde{c}} Q_{\text{L}(A)}{}^\text{C},
\qquad
    S_{\text{R}(A)} = \varepsilon_{\tilde{c}} S_{\text{L}(A)}{}^\text{C}.
\end{align}
Here $Q_{\text{L}(A)}{}^\text{C}$ and $S_{\text{L}(A)}{}^\text{C}$ are charge-conjugated fields. Using the mentioned properties of the generators, it can be proved readily that the right-handed spinors $Q_{\text{R}(A)}$ and $S_{\text{R}(A)}$ have all the same quantum numbers as the left-handed ones but the opposite-sign hypercharges:
\begin{align}
(Q^{j\underline j}_{\text{R}(A)})'&{}=Q^{j\underline j}_{\text{R}(A)} + \frac{i}{2} g_1 Y_{Q(A)} \theta
Q^{j\underline j}_{\text{R}(A)}+\frac{i}{2}g_2 \theta_a
\tau_a^{jk}Q^{k\underline j}_{\text{R}(A)}+\frac{i}{2}g_{\tilde{c}}\theta_{\underline a}\lambda_{\underline a}^{\underline j \underline k}
Q^{j\underline k}_{\text{R}(A)},
\\
(S^{\underline j}_{\text{R}(A)})'&{}=S^{\underline j}_{\text{R}(A)} + ig_1 Y_{S(A)} \theta
S^{\underline j}_{\text{R}(A)}+\frac{i}{2}g_{\tilde{c}}\theta_{\underline a}\lambda_{\underline a}^{\underline j \underline k}S^{\underline k}_{\text{R}(A)}.
\end{align}
Now it is easy see that we can form a doublet and a singlet of Dirac fields, $Q$ and $S$, simply setting $Y_{L(1)}=-Y_{L(2)}=Y_L$ and  $Y_{S(1)}=-Y_{S(2)}=Y_S$:
\begin{align}
Q=Q_{L(1)} + Q_{R(2)},
\qquad
S=S_{L(1)} + S_{R(2)}.
\end{align}
The above mentioned relations among hypercharges are also dictated by cancellation of gauge anomalies.

The three-flavor Lagrangian of H-quarks $Q$ and $S$ invariant under $G=G_\text{SM} \times \text{Sp}(2\chi_{\tilde{c}})$ reads
\begin{gather}\label{eq:LQS1}
    \mathscr{L} = i \bar Q \slashed{D} Q - m_Q \bar{Q} Q + i \bar S \slashed{D} S - m_S \bar{S} S,
\\
    D^\mu Q = \left[ \partial^\mu + \frac{i}{2} g_1 Y_Q B^\mu - \frac{i}{2} g_2 W_a^\mu \tau_a - \frac{i}2 g_{\tilde{c}} H^\mu_{\underline{a}} \lambda_{\underline{a}} \right] Q,
\\
    D^\mu S = \left[ \partial^\mu + i g_1 Y_S B^\mu - \frac{i}2 g_{\tilde{c}} H^\mu_{\underline{a}} \lambda_{\underline{a}} \right] S,
\end{gather}
where $H^\mu_{\underline{a}}$, $\underline{a} = 1 \dots \chi_{\tilde{c}} (2\chi_{\tilde{c}} + 1)$ are hypergluon fields. The electroweak interactions of H-quarks are chirally symmetric in this scheme, i.e.\ H-quarks are vectorlike, which is beneficial in view of electroweak precision constraints and makes mass terms gauge invariant. Note also that the Lagrangian can contain a direct Yukawa coupling of H-quarks with the SM Higgs doublet $\mathscr{H}$, if the H-quark hypercharges satisfy certain relations:
\begin{gather}\label{eq:LQHS}
    \delta\mathscr{L}_\text{int} = y_\text{L} \left( \bar{Q}_\text{L} \mathscr{H} \right) S_\text{R} + y_\text{R} \left( \bar{Q}_\text{R} \varepsilon \bar{\mathscr{H}} \right) S_\text{L} + \text{h.c.} \quad \text{ for } \frac{Y_Q}{2}-Y_S = +\frac12;
\\
    \delta\mathscr{L}_\text{int} = y_\text{L} \left( \bar{Q}_\text{L} \varepsilon \bar{\mathscr{H}} \right) S_\text{R} + y_\text{R} \left( \bar{Q}_\text{R} \mathscr{H} \right) S_\text{L} + \text{h.c.} \quad \text{ for } \frac{Y_Q}{2}-Y_S = -\frac12,
\\
    \delta\mathscr{L}_\text{int} = y_\text{L} \left( \bar{Q}_\text{L} \mathscr{H} \right) \varepsilon_{\tilde{c}} S^\text{C}{}_\text{R} + y_\text{R} \left( \bar{Q}_\text{R} \varepsilon \bar{\mathscr{H}} \right) \varepsilon_{\tilde{c}} S^\text{C}{}_\text{L} + \text{h.c.} \quad \text{ for } \frac{Y_Q}{2}+Y_S = +\frac12;
\\
    \delta\mathscr{L}_\text{int} = y_\text{L} \left( \bar{Q}_\text{L} \varepsilon \bar{\mathscr{H}} \right) \varepsilon_{\tilde{c}} S^\text{C}{}_\text{R} + y_\text{R} \left( \bar{Q}_\text{R} \mathscr{H} \right) \varepsilon_{\tilde{c}} S^\text{C}{}_\text{L} + \text{h.c.} \quad \text{ for } \frac{Y_Q}{2}+Y_S = -\frac12.
\end{gather}

Another consequence of H-quarks being in self-contragridient rep is that the kinetic term of the Lagrangian \eqref{eq:LQS1} can be written in terms of sextet of left-handed fields as follows:
\begin{gather}\label{eq:LP}
    \mathscr{L} = i \bar P_\text{L} \slashed{D} P_\text{L} ,
    \qquad P_\text{L} = \begin{pmatrix} Q_{L(1)} \\ Q_{L(2)} \\ S_{L(1)} \\ S_{L(2)} \end{pmatrix},
    \\\label{eq:covdcur}
    D^\mu P_\text{L} = \left[ \partial^\mu + i g_1 B^\mu \left( Y_Q \Sigma_Q + Y_S \Sigma_S \right) - \frac{i}{2} g_2 W_a^\mu \Sigma^a_W - \frac{i}2 g_{\tilde{c}} H^\mu_{\underline{a}} \lambda_{\underline{a}} \right] P_\text{L},
\\ \label{eq:SigmaQSW}
    \Sigma_Q = \frac12 \begin{pmatrix} 1 & 0 & 0 \\ 0 & -1 & 0 \\ 0 & 0 & 0 \end{pmatrix},
\qquad
    \Sigma_S = \begin{pmatrix} 0 & 0 & 0 \\ 0 & 0 & 0 \\ 0 & 0 & \tau_3 \end{pmatrix},
\qquad
    \Sigma^a_W = \begin{pmatrix} \tau_a & 0 & 0 \\ 0 & \tau_a & 0 \\ 0 & 0 & 0 \end{pmatrix}.
\end{gather}
In the limit of $g_1=g_2=0$ this Lagrangian is invariant under global SU(6) symmetry \cite{Pauli1957Conservation, Gursey1958Relation} that includes the chiral symmetry $\text{SU}(3)_\text{L} \times \text{SU}(3)_\text{R}$ as its subgroup:
\begin{gather}\label{eq:PLTP}
    P_\text{L} \to U P_\text{L}, \qquad U \in \text{SU}(6)
\end{gather}
It is broken explicitly by EW and Yukawa interactions, H-quark masses, and dynamically by H-quark condensate \cite{Vysotskii1985Spontaneous,Verbaarschot2004Supersymmetric}:
\begin{gather}\label{eq:LTQ}
    \langle \bar QQ + \bar SS \rangle = \frac12 \langle  \bar P_\text{L} M_0 P_\text{R} + \bar P_\text{R} M_0^\dagger P_\text{L} \rangle,
\qquad P_\text{R}  = \varepsilon_{\tilde{c}} P_\text{L}{}^\text{C},
    \qquad  M_0 = \begin{pmatrix} 0 & \varepsilon & 0\\ \varepsilon & 0 & 0 \\ 0 & 0 & \varepsilon \end{pmatrix},
\end{gather}
The condensate \eqref{eq:LTQ} is left invariant under transformations $U \in \text{SU}(6)$ that satisfy a condition
\begin{gather}\label{eq:symprel}
    U^\text{T} M_0 + M_0 U =0.
\end{gather}
They form Sp(6) subgroup of SU(6). Mass terms break the symmetry further to $\text{Sp(4)}\times\text{Sp}(2)$:
\begin{gather}\label{eq:massterm}
    \mathscr{L}_\text{masses} = -\frac12 \bar P_\text{L} M_0' P_\text{R} + \text{h.c.},
\qquad M'_0 = -M'_0{}^\text{T} = \begin{pmatrix} 0 & m_Q \varepsilon & 0\\ m_Q \varepsilon & 0 & 0 \\ 0 & 0 & m_S \varepsilon \end{pmatrix}.
\end{gather}


\section{\label{sec:LSM}Linear sigma model as an effective field theory of constituent H-quarks}

\subsection{Interactions of constituent H-quarks with composite scalars and intermediate bosons}

Let us now construct an effective Lagrangian of constituent H-quark interactions as a linear $\sigma$-model:
\begin{gather}\label{eq:ctq}
    \mathscr{L} = i \bar P_\text{L} \slashed{D} P_\text{L} -\sqrt2 \varkappa \left( \bar P_\text{L} M P_\text{R} + \bar P_\text{R} M^\dagger P_\text{L} \right),
\end{gather}
where $\varkappa$ is a coupling constant of hyperquark-hyperhadron interactions and the scalar field $M$ is a complex antisymmetric $6 \times 6$ matrix. The field $M$ transforms under the global symmetry SU(6) as follows:
\begin{gather}\label{eq:M_transf}
    M \to UMU^T, \qquad U \in \text{SU}(6) .
\end{gather}
For constituent H-quarks the electroweak group is embedded into unbroken Sp(6) subgroup of SU(6) in the same way as for the current ones \eqref{eq:covdcur}:
\begin{gather}
    D^\mu P_\text{L} = \left[ \partial^\mu + i g_1 B^\mu \left( Y_Q \Sigma_Q + Y_S \Sigma_S \right) - \frac{i}{2} g_2 W_a^\mu \Sigma^a_W \right] P_\text{L}.
\end{gather}

The field $M$ can be expanded in terms of 14 ``broken'' generators $\beta_a$ of the global symmetry group SU(6):
\begin{gather}
    M = \left[ \frac1{2\sqrt3} (A_0+iB_0) + (A_a+iB_a) \beta_a \right] M_0
\notag\\
    = \frac12 \begin{pmatrix} \bar{A}^Y\varepsilon & \left[ \frac{1}{\sqrt3} \sigma + \frac{1}{\sqrt6} f + \frac{1}{\sqrt2} a_a \tau_a \right] \varepsilon & K^\star \varepsilon \\
                                         \left[ \frac{1}{\sqrt3} \sigma + \frac{1}{\sqrt6} f - \frac{1}{\sqrt2} a_a \tau_a \right] \varepsilon & A \varepsilon & \varepsilon \bar{K}^\star \\
                                         K^{\star\dagger} \varepsilon & \varepsilon K^{\star\text{T}} & \frac{1}{\sqrt3} \left( \sigma - \sqrt2 f \right) \varepsilon
                \end{pmatrix}
\notag\\
        + \frac{i}2 \begin{pmatrix} \bar{B}^Y \varepsilon & \left[ \frac{1}{\sqrt3}\eta + \frac{1}{\sqrt6} \eta' + \frac{1}{\sqrt2} \pi_a \tau_a \right] \varepsilon & K \varepsilon \\
                                         \left[ \frac{1}{\sqrt3} \eta + \frac{1}{\sqrt6} \eta' - \frac{1}{\sqrt2} \pi_a \tau_a \right] \varepsilon & B \varepsilon & \varepsilon \bar{K} \\
                                         K^\dagger \varepsilon & \varepsilon K^\text{T} & \frac{1}{\sqrt3} \left( \eta - \sqrt2 \eta' \right) \varepsilon
                \end{pmatrix} ,
\end{gather}
\begin{gather}
    \sigma = A_0, \quad \eta = B_0, \quad
    f = A_6, \quad \eta' = B_6, \quad
    a_a = A_{a+2}, \quad \pi_a = B_{a+2}, \quad
    a=1,\,2,\,3,
    \notag\\
    A = \frac1{\sqrt2} ( A_1 + i A_2 ), \quad
    B = \frac1{\sqrt2} ( B_1 + i B_2 ),
    \notag\\
    K^\star = \frac12 \left[ A_{10} + i A_{14} + ( A_{6+a} + i A_{10+a} ) \tau_a \right],
    \notag\\
    K = \frac12 \left[ B_{10} + i B_{14} + ( B_{6+a} + i B_{10+a} ) \tau_a \right].
\notag\end{gather}
Explicit expressions for the generators $\beta_a$ are given in the appendix \ref{app:gens}. Bars over scalar fields denote simple complex conjugation.

Now the Lagrangian of constituent H-quarks \eqref{eq:ctq} can be put into the following form:
\begin{gather}
    \mathscr{L} = i \bar Q \slashed{D} Q + i \bar S \slashed{D} S - \varkappa u \left( \bar Q Q + \bar S S \right)
    \notag\\
    -\varkappa \bar{Q} \left[ \sigma' + \frac{1}{\sqrt3} f + i \left( \eta + \frac{1}{\sqrt3} \eta' \right) \gamma_5
     + \left( a_a + i \pi_a \gamma_5 \right) \tau_a
    \right]  Q
    \notag\\
    -\varkappa \bar{S} \left[ \sigma' - \frac{2}{\sqrt3} f + i \left( \eta - \frac{2}{\sqrt3} \eta' \right) \gamma_5 \right] S
    -\sqrt2 \varkappa \left[ \left( \bar Q \mathscr{K}^\star \right) S + i \left( \bar Q \mathscr{K} \right) \gamma_5 S + \text{h.c.} \right]
    \notag\\
    -\frac{\varkappa}{\sqrt2}  \left( A \bar Q \varepsilon \varepsilon_{\tilde{c}} Q^\text{C}
    +iB \bar Q \gamma_5 \varepsilon \varepsilon_{\tilde{c}} Q^\text{C} + \text{h.c.} \right)
    \notag\\
    -\sqrt2 \varkappa \left[ \left( \bar Q \mathscr{A} \right) \varepsilon_{\tilde{c}} S^\text{C}
                                     + i \left( \bar Q \mathscr{B} \right) \gamma_5 \varepsilon_{\tilde{c}} S^\text{C} + \text{h.c.} \right],
\end{gather}
where the singlet meson $\sigma$ has developed a v.e.v., $\sigma = u + \sigma'$, and we have defined $\text{SU}(2)_\text{L}$ doublets of H-mesons, $\mathscr{K}^\star$ and $\mathscr{K}$, and H-baryons (scalar diquarks), $\mathscr{A}$ and $\mathscr{B}$:
\begin{gather}
    \mathscr{K}^\star = \frac{1}{\sqrt2} \left( R_1 + i R_2 \right),
\qquad
    \mathscr{K} = \frac{1}{\sqrt2} \left( S_1 + i S_2 \right),
\\
    \mathscr{A} = \frac{1}{\sqrt2} \varepsilon \left( \bar R_1 + i \bar R_2 \right),
\qquad
    \mathscr{B} = \frac{1}{\sqrt2} \varepsilon \left( \bar S_1 + i \bar S_2 \right),
\\
    R_1 = \frac{1}{\sqrt2} \begin{pmatrix} A_{10} + i A_{13} \\ -A_{12} + i A_{11} \end{pmatrix},
\qquad
    R_2 = \frac{1}{\sqrt2} \begin{pmatrix} A_{14} - i A_{9} \\ A_{8} - i A_{7} \end{pmatrix},
\\
    S_1 = \frac{1}{\sqrt2} \begin{pmatrix} B_{10} + i B_{13} \\ -B_{12} + i B_{11} \end{pmatrix},
\qquad
    S_2 = \frac{1}{\sqrt2} \begin{pmatrix} B_{14} - i B_{9} \\ B_{8} - i B_{7} \end{pmatrix}.
\end{gather}
The covariant derivatives for H-quarks read
\begin{gather}
    D_\mu Q = \partial_\mu Q + \frac{i}2 g_1 Y_Q B_\mu Q - \frac{i}2 g_2 W_\mu^a \tau_a Q,
    \qquad
    D_\mu S = \partial_\mu S + i g_1 Y_S B_\mu S.
\end{gather}
The H-quark charges are $Q^U_\text{em}  = (Y_Q+1)/2$, $Q^D_\text{em} = (Y_Q-1)/2$, and $Q^S_\text{em} = Y_S$ (in units of the positron charge $e=|e|$).


\subsection{Interactions of H-hadrons with intermediate bosons}

The kinetic terms of the scalar and pseudoscalar fields in the model can be written as
\begin{align}\label{eq:LM}
    \mathscr{L} ={}& D_\mu \mathscr{H}^\dagger \cdot D^\mu \mathscr{H} + \Tr D_\mu M^\dagger \cdot D^\mu M
    \notag\\
    = {}&\frac12 \sum_\varphi D_\mu \varphi \cdot D^\mu \varphi  + \sum_\Phi D_\mu \Phi \cdot D^\mu \Phi
	 + D_\mu \bar A \cdot D^\mu A + D_\mu \bar B \cdot D^\mu B,
\end{align}
Here, $\varphi = h$, $h_a$, $\pi_a$, $a_a$, $\sigma$, $f$, $\eta$, $\eta'$ are all singlets and triplets, $\Phi = \mathscr{K}$, $\mathscr{K}^\star$, $\mathscr{A}$, $\mathscr{B}$ are doublets, and $\mathscr{H}$ is the fundamental (not composite) Higgs doublet of SM with conventionally defined transformation properties---the convariant derivative of $\mathscr{H}$ reads
\begin{gather}
    D_\mu \mathscr{H} = \left[ \partial_\mu + \frac{i}{2} g_1 B_\mu - \frac{i}{2} g_2 W_\mu^a \right] \mathscr{H},
\end{gather}
or, for components of $\mathscr{H}$,
\begin{gather}
   \mathscr{H} = \frac{1}{\sqrt2} \begin{pmatrix} h_2+ih_1 \\ h-ih_3\end{pmatrix},
\qquad
    D_\mu h
        = \partial_\mu h
             +\frac12 (g_1 \delta_3^a B_\mu +g_2 W_\mu^a ) h_a,
\\
    D_\mu h_a
        =\partial_\mu h_a
             -\frac12 (g_1 \delta_3^a B_\mu +g_2 W_\mu^a ) h
             -\frac12 e_{abc} (g_1 \delta_3^b B_\mu -g_2 W_\mu^b ) h_c.
\end{gather}

The electroweak group is embedded into Sp(6) as is seen in the form of the covariant derivative \eqref{eq:covdcur} of the left-handed H-quark sextet. Therefore, the covariant derivative for the scalar field $M$ can be readily written comparing transformation laws \eqref{eq:M_transf} and \eqref{eq:PLTP}:
\begin{align}
    D_\mu M ={}& \partial_\mu M
         + i Y_Q g_1 B_\mu (\Sigma_Q M + M \Sigma_Q^\text{T} )
         + i Y_S g_1 B_\mu (\Sigma_S M + M \Sigma_S^\text{T} )
\notag\\
         &- \frac{i}{2} g_2 W_\mu^a (\Sigma^a_W M + M \Sigma^{a\text{T}}_W{} )
\end{align}
where the matrices $\Sigma_Q$, $\Sigma_S$, $\Sigma^a_W$, $a=1,\,2,\,3$ are defined by Eq.~\eqref{eq:SigmaQSW}. Then, the covariant derivatives of H-baryons and H-mesons are as follows:
\begin{gather}\label{eq:dpi}
    D_\mu \pi_a = \partial_\mu \pi_a + g_2 e_{abc} W_\mu^b \pi_c ,
    \qquad
    D_\mu a_a = \partial_\mu a_a + g_2 e_{abc} W_\mu^b a_c,
    \\
    D_\mu A = \partial_\mu A + i g_1 Y_Q B_\mu A,
    \qquad
    D_\mu B = \partial_\mu B + i g_1 Y_Q B_\mu B,
    \\
    D_\mu \mathscr{K} = \left[ \partial_\mu + i g_1 \left( \frac{Y_Q}{2}-Y_S \right ) B_\mu - \frac{i}{2} g_2 W_\mu^a \tau^a \right] \mathscr{K},
	\\
	D_\mu \mathscr{K}^\star = D_\mu \mathscr{K} \bigr|_{\mathscr{K} \to \mathscr{K}^\star},
	\quad
	D_\mu \mathscr{A} = D_\mu \mathscr{K} \biggr|{}_{ \begin{subarray}{l} \mathscr{K} \to \mathscr{A} \\ Y_S \to -Y_S \end{subarray} },
	\quad
	D_\mu \mathscr{B} = D_\mu \mathscr{K} \biggr|{}_{ \begin{subarray}{l} \mathscr{K} \to \mathscr{B} \\ Y_S \to -Y_S \end{subarray} }.
\end{gather}

\begin{table}
\tbl{Quantum numbers of the lightest (pseudo)scalar H-hadrons and the corresponding H-quark currents in $Sp(2\chi_{\tilde{c}})$ model. $T$ is the weak isospin. $\tilde G$ denotes hyper-$G$-parity of a state (see Section \ref{sec:vars}). $\tilde B$ is the H-baryon number. $Q_\text{em}$ is the electric charge.}
{\begin{tabular}{ccccccccc}
\toprule
state & $$ & H-quark current & $$ & $T^{\tilde G}(J^{PC})$ & $$ & $\tilde B$ & $$ & $Q_\text{em}$ \\
\colrule
$\sigma$ & $$ & $\bar Q Q + \bar SS$ & $$ & $0^+(0^{++})$ & $$ & 0 & $$ & 0 \\
$f$ & $$ & $\bar Q Q -2 \bar SS$ & $$ & $0^+(0^{++})$ & $$ & 0 & $$ & 0 \\
$\eta$ & $$ & $i \left( \bar Q \gamma_5 Q + \bar S \gamma_5 S \right)$ & $$ & $0^+(0^{-+})$ & $$ & 0 & $$ & 0 \\
$\eta'$ & $$ & $i \left( \bar Q \gamma_5 Q - 2 \bar S \gamma_5 S \right)$ & $$ & $0^+(0^{-+})$ & $$ & 0 & $$ & 0 \\
$ a_k$ & $$ & $\bar Q \tau_k Q$ & $$ & $1^-(0^{++})$ & $$ & 0 & $$ & $\pm 1$, 0 \\
$\pi_k$ & $$ & $i \bar Q \gamma_5 \tau_k Q$ & $$ & $1^-(0^{-+})$ & $$ & 0 & $$ & $\pm 1$, 0 \\
$A$ & $$ & $\bar Q^\text{C} \varepsilon \varepsilon_{\tilde{c}} Q$ & $$ & $0^{\hphantom{+}}(0^{-\hphantom{+}})$ & $$ & 1 & $$ & $Y_Q$ \\
$B$ & $$ & $i \bar Q^\text{C}  \varepsilon \varepsilon_{\tilde{c}} \gamma_5 Q$ & $$ & $0^{\hphantom{+}}(0^{+\hphantom{+}})$ & $$ & 1 & $$ & $Y_Q$ \\
$\mathscr{K}^\star$ & $$ & $\bar S Q$ & $$ & $\frac12^{\hphantom{+}}(0^{+\hphantom{+}})$ & $$ & 0 & $$ & $Y_Q/2-Y_S \pm 1/2$ \\
$\mathscr{K}$ & $$ & $i \bar S \gamma_5 Q$ & $$ & $\frac12^{\hphantom{+}}(0^{-\hphantom{+}})$ & $$ & 0 & $$ & $Y_Q/2-Y_S \pm 1/2$ \\
$\mathscr{A}$ & $$ & $\bar S^\text{C} \varepsilon_{\tilde{c}} Q$ & $$ & $\frac12^{\hphantom{+}}(0^{-\hphantom{+}})$ & $$ & 1 & $$ & $Y_Q/2+Y_S \pm 1/2$ \\
$\mathscr{B}$ & $$ & $i \bar S^\text{C} \varepsilon_{\tilde{c}} \gamma_5 Q$ & $$ & $\frac12^{\hphantom{+}}(0^{+\hphantom{+}})$ & $$ & 1 & $$ & $Y_Q/2+Y_S \pm 1/2$ \\
\botrule
\end{tabular} \label{tab:H-hadrons} }
\end{table}


\subsection{Self-interactions and masses of the (pseudo)scalar fields}

The lowest-dimension invariants contributing to the self-interaction Lagrangian of (pseudo)scalars are:
\begin{gather}\label{eq:Uinvs}
    I_0 = \mathscr{H}^\dagger \mathscr{H},
    \quad
    I_1 = \Tr \left( M^\dagger M \right),
    \quad
    I_2 = \Tr \left[ \left( M^\dagger M \right)^2 \right],
    \quad
    I_3 = \mathop{\mathrm{Re}} \mathop{\mathrm{Pf}} M,
\end{gather}
where $\mathop{\mathrm{Pf}} M = \frac1{2^3 3!} \varepsilon_{abcdef} M_{ab} M_{cd} M_{ef}$ is the Pfaffian of $M$, $\varepsilon_{abcdef} $ is the 6-dimensional Levi-Civita symbol ($\varepsilon_{123456}=+1$). All the invariants \eqref{eq:Uinvs} are CP even. The potential of the Higgs and scalar H-hadrons reads
\begin{gather}\label{eq:U}
    U = - \mu_0^2 I_0 - \mu_1^2 I_1 + \lambda_{01} I_0 I_1 + \lambda_{11} I_1^2 + \lambda_2 I_2 + \lambda_3 I_3.
\end{gather}
Here we restrict ourselves to considering only renormalizable piece of the potential.

Now we can write tadpole equations for v.e.v.'s $v = \langle h \rangle \neq 0$ and $u = \langle \sigma \rangle \neq 0$ (here, we consider the case of vanishing Yukawa couplings $y_\text{L} = y_\text{R} = 0$):
\begin{gather}\label{eq:tadpoles} 
    \mu_0^2 = \lambda_{0} v^2 + \frac12 \lambda_{01} u^2,
    \quad
    \mu_1^2 = \left( \lambda_{1} + \frac16 \lambda_2 \right) u^2 + \frac12 \lambda_{01} v^2 - \frac{\sqrt{3}}{24} \lambda_3 u + \frac{\zeta \langle \bar Q Q + \bar S S \rangle}{u}.
\end{gather}

Vacuum stability is ensured by the following inequalities:
\begin{gather}\label{eq:vacineq} 
    \Lambda_{1} = \lambda_{1} + \frac16 \lambda_2 - \frac{\sqrt{3}}{48 u} \lambda_3 - \frac{\zeta \langle \bar Q Q + \bar S S \rangle}{2u^3} > 0, \qquad \lambda_{0} > 0,
    \qquad
    4 \lambda_{0} \Lambda_{1} - \lambda_{01}^2 > 0.
\end{gather}
In deriving \eqref{eq:tadpoles} and \eqref{eq:vacineq} the explicit breaking of the $SU(6)$ global symmetry have been represented by a Lagrangian term $\mathscr{L}_\text{SB} = -\zeta \langle \bar Q Q + \bar S S \rangle (u+\sigma')$, where $\zeta$ is a parameter proportional to the current mass $m_Q$ of the H-quarks. We should note that the effects of the explicit breaking can be taken in account by
different non-invariant terms \cite{1979PhRvC..19.1965C,2000PhRvC..61b5205D,1996NuPhA.603..239D}, but  the most common one, which is sometimes reffered to as ``standard breaking'', is a tadpole-like $\sigma$ term (see \refcite{1969RvMP...41..531G,2013PhRvD..87a4011P}, for example).

Masses of (pseudo)scalar fields:
\begin{gather}
    m_{\sigma,H}^2 = \lambda_{0} v^2 + \Lambda_{1} u^2 \pm \sqrt{(\lambda_{0} v^2 - \Lambda_{1} u^2)^2 + \lambda_{01}^2 v^2 u^2},
    \\
    m_\eta^2 = \frac{\sqrt{3}}{8} \lambda_{3} u  -\frac{\zeta \langle \bar Q Q + \bar S S \rangle}{u},
    \;\;
    m_\pi^2 = m_{\eta'}^2 = m_B^2 = m_\mathscr{K}^2 = m_\mathscr{B}^2 = -\frac{\zeta \langle \bar Q Q + \bar S S \rangle}{u},
    \\
     m_a^2 = m_{f}^2 = m_A^2 = m_\mathscr{K^\star}^2 = m_\mathscr{A}^2  = \frac{\sqrt{3}}{12} \lambda_{3} u + \frac13 \lambda_{2} u^2  -\frac{\zeta \langle \bar Q Q + \bar S S \rangle}{u}.
\end{gather}
The mixing of the scalars is as follows:
\begin{gather}
    h = \cos\theta_s H - \sin\theta_s \sigma,
    \qquad
    \sigma' = \sin\theta_s H + \cos\theta_s \sigma,
    \\[3mm]
    \tan2\theta_s = \frac{\lambda_{01} v u }{\lambda_{0} v^2 - \Lambda_{1} u^2},
    \qquad
    \sgn \sin\theta_s = -\sgn \lambda_{01},
\end{gather}
where $h$ and $\sigma'$ are the fields being mixed, while $H$ and $\sigma$ are physical ones.


\subsection{\label{sec:vars}Accidental symmetries}

For some particular choices of H-quark hypercharges the renormalizable Lagrangian of the model under consideration can be invariant under a number of additional symmetries. It is easy to see that several options are available, of which two are as follows.

(1)  \underline{$Y_Q = Y_S = 0$.} In the case of zero H-quark hypercharges the singlet H-baryons are neutral, while the states from H-baryon doublets carry electric charges of $\pm 1/2$. The singlet H-quark $S$ does not participate in EW interactions. The Lagrangian of current H-quarks is invariant under a generalization of $G$-parity of QCD that we dub as hyper $G$-parity \cite{2010PhRvD..82k1701B,Antipin:2015xia}:
\begin{align}\label{eq:HGconjugation}
	Q^{\tilde{\text{G}}} = \varepsilon \varepsilon_{\tilde{c}} Q^\text{C},
	\qquad
	S^{\tilde{\text{G}}} = \varepsilon_{\tilde{c}} S^\text{C}.
\end{align}
H-gluons and all SM degrees of freedom are postulated to be invariant under $\tilde G$-parity. As a result, the neutral H-pion $\pi^0$ is stable from decaying by the symmetry as the lightest $\tilde G$-odd state in the model. Besides, the Lagrangian is invariant under two global U(1) groups so that numbers of $Q=U,\,D$ and $S$ H-quarks are conserved. This forbids decays of neutral H-baryon $B$ and the lightest state in doublet $\mathscr{B}$. Therefore, we have three stable particles in this case---the neutral H-pion and two H-baryons, neutral $B$ and the lightest of two $\mathscr{B}$ states with a charge of $\pm 1/2$.

(2) \underline{$Y_Q = 0$, $Y_S = \pm 1/2$.} In terms of H-hadron charges this case is the most similar to QCD---H-kaons and doublet H-baryons have charges $\pm 1$ and 0, while singlet H-baryons are still neutral as in the first case. The Yukawa couplings of H-quarks with elementary Higgs is permitted by the gauge symmetries, which can lead to a mixing of the SM Higgs doublet and H-kaons. The invariance under global U(1) groups and $\tilde G$-conjugation \eqref{eq:HGconjugation}, however, is in general broken by the Yukawa couplings and the hypercharge of the singlet H-quark. Therefore, this variant of the model does not have stable states, except possibly for the lightest H-baryon.


\section{\label{sec:pt}Physical Lagrangian---interactions of new fields with gauge bosons}

Here, we give interactions of new physical particles with standard gauge bosons for the case of arbitrary hypercharges $Y_Q$ and $Y_S$. Corresponding vertices will be used in calculations of oblique parameters. The H-quark interactions with the EW bosons are vectorlike and their Lagrangian reads
\begin{align}\label{1}
\mathscr L=&{}-\frac{1}{2}e(Y_Q+1) A_{\mu} \bar{U}\gamma^{\mu}U-  \frac{1}{2}e(Y_Q-1) A_{\mu}\bar{D}\gamma^{\mu}D- eY_S A_{\mu}\bar{S}\gamma^{\mu}S\notag\\
      &{}+\frac{1}{2}(g_2 c_w-g_1s_wY_Q) Z_{\mu}\bar{U}\gamma^{\mu}U-\frac{1}{2}(g_2 c_w+g_1s_wY_Q)Z_{\mu}\bar{D}\gamma^{\mu}D
\notag\\
     & {}- g_1 s_w Y_S Z_{\mu}\bar{S}\gamma^{\mu}S+\frac{1}{\sqrt{2}}g_2 W^+_{\mu}\bar{U}\gamma^{\mu}D+\frac{1}{\sqrt{2}}g_2 W^-_{\mu}\bar{D}\gamma^{\mu}U,
\end{align}
where $e=g_1 c_w=g_2 s_w$ and $s_w=\sin\theta_w$ is the sine of the Weinberg angle.

Interactions of (pseudo)scalar H-mesons with photons and intermediate bosons are described by the following Lagrangian:
\begin{align}\label{2}
\mathscr L ={}&
 i g_2 c_w W_+^\mu \left( \pi^0  \pi^-_{,\mu}- \pi^- \pi^0_{,\mu} \right) +  i g_2 ( c_w Z^\mu + s_w A^\mu ) \pi^- \pi^+_{,\mu}
\notag\\
&{} + \frac12 g_2^2  \pi^+ \pi^- ( c_w Z^\mu + s_w A^\mu )^2
- g_2^2 \pi^0 ( c_w Z^\mu + s_w A^\mu ) \pi^+ W^-_\mu
 \notag\\ 
&{} + \frac12 g_2^2 \left( \pi_0^2+\pi^- \pi^+ \right) W^+_\mu W_-^\mu -\frac12 g_2^2 \pi_+^2 W^-_\mu W_-^\mu  + \text{h.c.} +(\pi \to  a)  .
\end{align}
In the above Lagrangian the last term means that the interactions of the triplet scalar H-mesons $ a$ have the same couplings and vertices as the interactions of the H-pions do. Interactions of gauge bosons with doublets $K=(K_1, K_2)$ and $K^\star=(K^\star_1, K^\star_2)$ are as follows:
\begin{align}\label{3}
\mathscr L={}&\sum_{k=1,2} \left[ i (Q_k A^{\mu} +G_k Z^{\mu})(\bar{K}_k\partial_{\mu}K_k-K_k\partial_{\mu}\bar{K}_k) + (Q_k A^{\mu} +G_k Z^{\mu})^2 \bar{K}_k K_k \right] \notag\\
          &+\biggl[ \frac{i}{\sqrt{2}}g_2 W^{\mu}_+(\bar{K}_1\partial_{\mu}K_2- \partial_{\mu} \bar{K}_1 K_2 )
			+\sqrt{2}q g_2 (c_w A^{\mu}-s_w Z^{\mu}) W^+_{\mu}\bar{K}_1 K_2 +\text{h.c.} \biggr]
          \notag\\
          &+\frac{1}{2}g^2_2 W^-_{\mu}W^{\mu +} (\bar{K}_1 K_1 +\bar{K}_2 K_2) +(K\to K^\star),
\end{align}
where the coupling constants $Q_i$, $G_i$, and $q$ are defined as follows:
\begin{equation}\label{4}
Q_{1,2}= -e\left(\frac{Y_Q}{2}-Y_S \pm \frac{1}{2} \right),
\quad
G_{1,2}=-qs_w\pm \frac{1}{2}g_2 c_w,
\quad
q=g_1 \left(\frac{Y_Q}{2}-Y_S \right).
\end{equation}
Interactions of gauge bosons with doublets $\mathscr A$ and $\mathscr B$ can be obtained from Eq.~(\ref{3}) by a substitution $Y_S \to -Y_S$, $K^\star \to \mathscr A$, $K \to \mathscr B$.

Now, we check that the model under consideration does not contradict precision EW constraints on effects of new physics. In particular, we consider the constraints on oblique corrections which are described by Peskin--Takeuchi (PT) parameters  \cite{PhysRevD.46.381}. We use the definitions of PT oblique parameters  from Ref.~\refcite{1994PhLB..326..276B}. These parameters are expressed in terms of vector bosons transversal polarizations $\Pi_{ab}(p^2)$, where $a,b=W, Z, \gamma$. Here, we use the definition $\Pi_{\mu\nu}^{ab}(p^2)=p_{\mu}p_{\nu}P_{ab}(p^2) + g_{\mu\nu}\Pi_{ab}(p^2)$, where $\Pi^{ab}_{\mu\nu}(p^2)$ is the polarization operator as a function of external momentum $p$. In the case under consideration, $\Pi_{ab}(0)=0$ and $\Pi'_{ab}(0) \sim \ln(M^2/\mu^2)-1/\bar{\epsilon}$. All divergencies and logarithms of the regularization parameter $\mu$ cancel out in the final expressions for oblique parameters. In the physically interesting case $Y_Q=0$ and $Y_S=-1/2$ the parameters can be represented in a simple form
\begin{align}\label{5}
S={}&\frac{2}{3\pi}(2c^4_w +s^4_w) \left[\chi_{\tilde{c}} F\left(\frac{M^2}{M^2_Z}\right)+\Phi\left(\frac{m^2}{M^2_Z}\right) \right],
\qquad
T=0,\notag\\
U={}&\frac{4}{3\pi}\left[\chi_{\tilde{c}} F\left(\frac{M^2}{M^2_W}\right)+\Phi\left(\frac{m^2}{M^2_W}\right)\right]-S.
\end{align}
In Eq.~(\ref{5}) $M=M_Q$, $m=m_{\tilde{\pi}}=m_K=m_{\mathscr B}$ and we assume $M_S=M_Q$ for simplicity. The functions $F(x)$ and $\Phi(x)$ are defined as follows:
\begin{align}\label{7}
F(x)&{}=-\frac13+(1+2x)f(x),
\qquad
\Phi(x)=\frac13+\frac12(1-4x)f(x),
\notag\\
f(x)&{}=2 \left( 1-\sqrt{4x-1}\arctan\frac{1}{\sqrt{4x-1}} \right).
\end{align}
From Eqs.~(\ref{5}) and \eqref{7} it follows that $S,\, U\to 0$ when $M^2/M^2_Z \to \infty$. Besides, $S<0.1$ for $M \approx 100$ GeV in the cases $\chi_{\tilde{c}}=2,4$. This value sets a lower bound on constituent masses of new quarks, while a bound for the masses of pseudo-goldstones is significantly less restrictive. In the case of heavy H-quarks, $M>500$ GeV the values of the parameters are $S<0.01$ and $U<0.001$. These values are smaller than the current experimental limits \cite{PhysRevD.98.030001}: $S=0.00^{+0.11}_{-0.10}$, $U=0.08 \pm 0.11$, $T=0.02^{+0.11}_{-0.12}$, that is the scenario under consideration satisfies the restrictions on indirect manifestations of new heavy  fermions. The additional parameters $V$, $W$, $X$ describe the contributions of new fermions with masses close to EW scale. In the case $Y_S=0$ the expression for oblique parameters can be derived from Eq.~(\ref{5}) by eliminating $s^4_w$ in the first equation. This does not affect the situation noticeably.


\section{Conclusion}

We have considered a hypercolor extension of SM with three flavors of Dirac H-quarks in a pseudoreal representation of an additional symplectic H-color group, i.e.\ the total symmetry group of the model is $G=G_\text{SM} \times \text{Sp}(2\chi_{\tilde{c}})$, $\chi_{\tilde{c}} \geqslant 1$. The model retains the elementary Higgs doublet of SM as one of its components. With respect to EW subgroup of $G_\text{SM}$ the original H-quarks are introduced as two left-handed doublets and two left-handed singlets. If the hypercharges of two doublets as well as singlets are of opposite signs one to another, the model is free of gauge anomalies and three flavors of the Dirac H-quarks are formed. The Dirac mass terms of H-quarks are allowed by the symmetry $G$ and the values of H-quark masses are assumed to lie lower than the scale of dynamical breaking of the global symmetry of the H-quark sector. The model possesses the global symmetry group SU(6) that is broken spontaneously to its subgroup Sp(6). There is also a possibility of an explicit symmetry breaking by H-quark masses and direct Yukawa couplings of H-quarks to the elementary Higgs doublet that are permitted by the symmetry $G$ for some values of H-quark hypercharges. Eventually, fourteen pseudo-Nambu--Goldstone states emerge---octet of H-mesons (there is a SU(3) subgroup of Sp(6)), two doublets and two singlets of scalar H-baryons. Depending on the choice of H-quark hypercharges, the model can benefit from accidental symmetries protecting some of the lightest states from decaying.

A linear sigma model has been constructed as an effective theory of constituent H-quark and H-hadron interactions. In addition to 14 pNG states the sigma model contains also a singlet sigma H-meson and 15 chiral partners of these H-hadrons. We have calculated contributions of these states to PT parameters. The experimental constraints on the parameters are satisfied for any reasonable values of H-hadron and constituent H-quark masses.


\appendix

\section{\label{app:gens}Algebras su(6) and sp(6)}

The generators $\Lambda_\alpha$, $\alpha=1,\dots,\,34$ of SU(6) group form a basis of all traceless hermitian matrices normalized so that $\Tr \Lambda_\alpha \Lambda_\beta = \frac12 \delta_{\alpha\beta}$.
Below we tabulate two subsets of the generators $\Lambda_\alpha$: $\Sigma_{\dot\alpha}$, $\dot\alpha=1,\dots,\,21$ and $\beta_\alpha$, $\alpha=1,\dots,\,14$. The first subset $\Sigma_{\dot\alpha}$ is a sub-algebra $\text{sp}(6) \subset \text{su}(6)$, i.e. $\Sigma_{\dot\alpha}$ satisfy a relation
$
    \Sigma_{\dot\alpha}^T M_0 + M_0 \Sigma_{\dot\alpha} =0,
$
where the antisymmetric matrix $M_0$ is chosen as in Eq.~\eqref{eq:LTQ}.
The rest of the generators, $\beta_a$, belong to the coset SU(6)/Sp(6) and satisfy a relation
$
    \beta_\alpha^T M_0 = M_0 \beta_\alpha.
$
The explicit expressions for the generators are as follows:
\begin{gather*}
    \Sigma_{a} = \frac{1}{2\sqrt2} \begin{pmatrix} 0 & \tau_a & 0 \\ \tau_a & 0 & 0 \\ 0 & 0 & 0 \end{pmatrix},
    \quad
    \Sigma_{3+a} = \frac{i}{2\sqrt2} \begin{pmatrix} 0 & \tau_a & 0 \\ -\tau_a & 0 & 0 \\ 0 & 0 & 0 \end{pmatrix},
    \quad
    \Sigma_{6+a} = \frac{1}{2} \begin{pmatrix} 0 & 0 & 0 \\ 0 & 0 & 0 \\ 0 & 0 & \tau_a \end{pmatrix},
\\
    \Sigma_{9+a} = \frac{1}{2\sqrt2} \begin{pmatrix} \tau_a & 0 & 0 \\ 0 & \tau_a & 0 \\ 0 & 0 & 0 \end{pmatrix},
    \quad
    \Sigma_{12+a} = \frac{1}{4} \begin{pmatrix} 0 & 0 & \tau_a \\ 0 & 0 & \tau_a \\ \tau_a & \tau_a & 0 \end{pmatrix},
    \quad
    \Sigma_{15+a} = \frac{i}{4} \begin{pmatrix} 0 & 0 & \tau_a \\ 0 & 0 & -\tau_a \\ -\tau_a & \tau_a & 0 \end{pmatrix},
\\
    \Sigma_{19} = \frac{1}{2\sqrt2} \begin{pmatrix} 1 & 0 & 0 \\ 0 & -1 & 0 \\ 0 & 0 & 0 \end{pmatrix},
    \quad
    \Sigma_{20} = \frac{1}{4} \begin{pmatrix} 0 & 0 & 1 \\ 0 & 0 & -1 \\ 1 & -1 & 0 \end{pmatrix},
    \quad
    \Sigma_{21} = \frac{i}{4} \begin{pmatrix} 0 & 0 & 1 \\ 0 & 0 &1 \\ -1 & -1 & 0 \end{pmatrix},
\end{gather*}

\begin{gather*}
    \beta_1 = \frac{1}{2\sqrt2} \begin{pmatrix} 0 & 1 & 0 \\ 1 & 0 & 0 \\ 0 & 0 & 0 \end{pmatrix},
    \qquad
    \beta_2 = \frac{i}{2\sqrt2} \begin{pmatrix} 0 & -1 & 0 \\ 1 & 0 & 0 \\ 0 & 0 & 0 \end{pmatrix},
    \qquad
    \beta_{2+a} = \frac{1}{2\sqrt2} \begin{pmatrix} \tau_a & 0 & 0 \\ 0 & -\tau_a & 0 \\ 0 & 0 & 0 \end{pmatrix},
\end{gather*}
\begin{gather*}
    \beta_{6} = \frac{1}{2\sqrt6} \begin{pmatrix} 1 & 0 & 0 \\ 0 & 1 & 0 \\ 0 & 0 & -2 \end{pmatrix},
    \qquad
    \beta_{6+a} = \frac{1}{4} \begin{pmatrix} 0 & 0 & \tau_a \\ 0 & 0 & -\tau_a \\ \tau_a & -\tau_a & 0 \end{pmatrix},
    \qquad
    \beta_{10} = \frac{1}{4} \begin{pmatrix} 0 & 0 & 1 \\ 0 & 0 &1 \\ 1 & 1 & 0 \end{pmatrix},
\\  \beta_{10+a} = \frac{i}{4} \begin{pmatrix} 0 & 0 & \tau_a \\ 0 & 0 & \tau_a \\ -\tau_a & -\tau_a & 0 \end{pmatrix},
    \qquad
    \beta_{14} = \frac{i}{4} \begin{pmatrix} 0 & 0 & 1 \\ 0 & 0 & -1 \\ -1 & 1 & 0 \end{pmatrix},
\end{gather*}
where $\tau_a$, $a=1,\,2,\,3$ are the Pauli matrices.


\section*{Acknowledgments}

The work was supported by Russian Scientific Foundation (RSCF) Grant \textnumero 18-12-00213.


\bibliographystyle{ws-ijmpd}
\bibliography{ijmpd_hc_18}

\end{document}